\documentclass[12pt]{article}
\usepackage{natbib,amsmath,amsfonts,fullpage,graphicx,setspace,amsthm}
\usepackage[colorlinks,linkcolor=black,citecolor=blue,urlcolor=black]{hyperref}

\usepackage{times}
\usepackage{bm}

\usepackage[plain,noend]{algorithm2e}

\newcommand\cites[1]{\citeauthor{#1}'s\ (\citeyear{#1})}

\usepackage{graphicx}                
\usepackage{amsmath,amssymb,bm} 
\usepackage{listings,multirow}
\usepackage{bbm}

\graphicspath{{figures/}} 

\newtheorem{theorem}{Theorem}
\newtheorem{corollary}{Corollary}
\newtheorem{definition}{Definition}

\begin{document}
\title{Diagnosing missing always at random in multivariate data\thanks{We thank Fabrizia Meall for her useful insights.}}
 \author{Iavor I. Bojinov, Natesh S. Pillai, and Donald B. Rubin \\
 \textit{Harvard University}}

\maketitle

\begin{abstract}
    Models for analyzing multivariate data sets with missing values require strong, often unassessable, assumptions. The most common of these is that the mechanism that created the missing data is ignorable - a twofold assumption dependent on the mode of inference. The first part, which is the focus here, under the Bayesian and direct-likelihood paradigms, requires that the missing data are missing at random; in contrast, the frequentist-likelihood paradigm demands that the missing data mechanism always produces missing at random data, a condition known as missing always at random. Under certain regularity conditions, assuming missing always at random leads to an assumption that can be tested using the observed data alone namely, the missing data indicators only depend on fully observed variables. Here, we propose three different diagnostic tests that not only indicate when this assumption is incorrect but also suggest which variables are the most likely culprits. Although missing always at random is not a necessary condition to ensure validity under the Bayesian and direct-likelihood paradigms, it is sufficient, and evidence for its violation should encourage the careful statistician to conduct targeted sensitivity analyses.
\end{abstract}


\section{Introduction}
    When conducting statistical analyses of data sets with missing values, researchers have to make assumptions that can not be assessed using the observed data alone, referred to as unassessable assumptions by \cite{liublinska2014sensitivity}. \cites{rubin1976inference} seminal paper was the first to formalize these assumptions by explicitly considering the missing data indicators as random variables as well as providing the weakest sufficient conditions that lead to correct inference about the parameter that governs the distribution of the data when ignoring the mechanism that created the missing data. \cite{rubin1976inference} showed that it is appropriate to ignore the missingness mechanism (i.e., the conditional distribution of the missing data indicators given the missing and the observed data) when using direct-likelihood \citep{fisher1956statistical,edwards1984likelihood} or Bayesian inference for the parameter governing the distribution of the data, $\theta$, when the missing data are missing at random (MAR) and parameter of the missingness mechanism, $\phi$, and $\theta$ are distinct. The missing data are MAR if the missingness mechanism evaluated at the observed missing data pattern and the observed data, considered as a function of the missing data and $\phi$, takes the same value for all possible values of the missing data and the parameter $\phi$. The definition of MAR depends on the missing data, and it is, therefore, generally unassessable as its validity cannot be falsified using the observed data alone. The much stronger assumption of missing completely at random (MCAR), which is sufficient, although not necessary, to ignore the missing data mechanism for valid frequentist inference, however, can be assessed \citep{little1988test,park1993test}. This is true because MCAR further imposes that, for all values of $\phi$, the missingness mechanism, evaluated at the observed missing data pattern, takes the same value for all values of the observed data as well as the missing data \citep{marini1980maximum}.

    Subsequently, there have been many misinterpretations of the conditions provided in \cite{rubin1976inference} (e.g., \cite{lu2004missing} used MAR to mean that the missingness mechanism always produces MAR data sets, and \cite{fitzmaurice2012applied} interpreted MAR as a conditional independence statement). \cite{seaman2013meant} and \cite{bimetrkia2015Mealli} are two recent papers that clarify the situation by explaining the difference between the data being missing at random and the missingness mechanism always producing missing at random data. This distinction is critical for understanding when ignoring the missingness mechanism yields valid statements from different inferential perspectives, and what assumptions can be assessed using the observed data alone and under what conditions.

    Although the MCAR assumption is sufficient for valid frequentist inference, certain aspects of likelihood-based frequentist inference (that is, using maximum likelihood estimates and the observed information matrix to measure their precision) are asymptotically valid when the missingness mechanism always produces MAR data sets \citep{molenberghslikelihood,little2002statistical}. \cite{seaman2013meant} referred to this type of missingness mechanism as ``everywhere missing at random,'' whereas \cite{bimetrkia2015Mealli} referred to it as ``missing always at random''; we choose to follow the latter suggestion as the word ``everywhere'' in probability and statistics has a different mathematical meaning, which is not reflected in its use in this context. A missing always at random (MAAR) missingness mechanism always produces MAR data, but not all MAR data sets are generated from a MAAR mechanism. Therefore, assuming MAAR and parameter distinctness is a sufficient condition for ignoring the missing data mechanism when conducting direct-likelihood, frequentist likelihood, or Bayesian inference. With Bayesian inference, the distinct parameters assumption requires independent prior distributions on $\phi$ and $\theta$. 

    \cite{bimetrkia2015Mealli} also formalized the general intuition that having more fully observed covariates makes the MAAR assumption more plausible; we restate the relevant theorem without proof in Section \ref{sec:Notation}. Utilizing this result, we derive two corollaries that we use to construct three different diagnostic tests for diagnosing the plausibility of MAAR. 
    Violating MAAR does not imply that Bayesian or direct-likelihood inference conducted by ignoring the missingness mechanism are invalid; however, data sets that satisfy MAR but not MAAR are ones where the resulting inference are more sensitive to model misspecification, in the sense first explored in \cite{rosenbaum1984sensitivity}. Each of our diagnostic tests identifies the variables that are more likely to break the MAAR assumption by looking for conditional dependencies between the missing data indicators and the variables with missing values, given the fully observed variables. Focusing on these variables can lead to better, more targeted, sensitivity analyses because these variables are likely creators of the violation of MAR. The diagnostic tests we propose should not be treated as formal hypothesis tests, but rather instead treated like diagnostic tools such as the ones presented in \cite{potthoff2006can,kobi2008diagnostics}; and \cite{bondarenko2016graphical}.

   In Section \ref{sec:Notation} we provide the notation, definitions and the main results that will be used throughout the paper. In Section \ref{sec:Methods} we introduce our three diagnostic tests and explain how to apply each of them to a data set. In Section \ref{sec:Simulation} we conduct a simulation study to analyze the frequentist operating characteristics of our proposed tests. In Section \ref{sec:Conclusion} we provide concluding comments.  

\section{Notation and definitions}\label{sec:Notation}

    Let $Y$ be the complete data matrix, arising from distribution $p(Y\mid \theta)$, with entries $Y_{i,j}$ corresponding to the (possibly missing) response of unit $i=1,\dots,N$ to variable $j=1,\dots, J$. Define $Y_{\cdot j} = (Y_{1,j} ,\dots , Y_{N,j})^T$ to be the random column vector of the unit responses for variable $j$, and let $Y_{i \cdot} = (Y_{i,1},\dots, Y_{i,J})$ be the random row vector of unit $i$'s responses. Define $R$ to be the response indicator matrix such that $R_{i,j} = 1$ if $Y_{i,j}$ is observed and 0 if $Y_{i,j}$ is missing. The missingness mechanism is then the conditional distribution of $R$ given $Y$ indexed by a parameter $\phi$, denoted by $p(R\mid Y,\phi)$. The response indicators $R_{\cdot j}=(R_{1,j},\dots,R_{N,j})$ partition $Y_{\cdot j}$ into two sets $\mathcal{S}(R_{\cdot j}, Y_{\cdot j}) = \{Y_{i,j}\ \text{such that} \ R_{i,j} = 1\}$ the observed values of $Y_{\cdot j}$ and $\mathcal{S}(1-R_{\cdot j}, Y_{\cdot j}) = \{Y_{i,j}\ \text{such that} \ R_{i,j} = 0\}$ the missing values of $Y_{\cdot j}$. In general $\mathcal{S}(R_{\cdot k}, Y_{\cdot j}) = \{Y_{i,j}\ \text{such that} \ R_{i,k} = 1\}$ will denote the set of $Y_{\cdot j}$ for which $R_{i,k} = 1$. The concatenation of $\mathcal{S}(R_{\cdot j}, Y_{\cdot j})$ will be written as $\mathcal{S}(R, Y)$ rather than the usual $Y_\text{obs}$ or $Y_{(1)}$. We have purposefully kept our notation very general so that we can readily consider the partition that a missingness indicator induces on all of the columns on $Y$.  Throughout most of the paper, we will assume that only $J^*<J$ of the columns of $Y$ can have missing values, then the matrix $Y_{\cdot J^*+1:J} = (Y_{\cdot J^*+1}, \dots, Y_{\cdot J})$ will always be fully observed, and $R$ will have $J^*$ columns. A generic value of $R$ and $Y$ will be denoted by $r$ and $ y$ respectively, whereas the realized values will be indicated by $ \tilde r$ and $\tilde y$, respectively.

    We now provide the formal definitions of MAR and MAAR, and then state the key theorem from \cite{bimetrkia2015Mealli}.
    \begin{definition}\label{D:MAAR}
        The missing data are \emph{missing at random} (MAR) if
        \begin{equation}\label{E:DMAR}
                p(R= \tilde r\mid Y = \tilde y, \phi) =p(R= \tilde r\mid Y = \tilde y', \phi),
            \end{equation}
        for all  $\phi$, and all $\tilde y$ and $\tilde y'$ such that $\mathcal{S}(\tilde r, \tilde y) = \mathcal{S}(\tilde r, \tilde y')$. \\ The missing data are \emph{missing always at random} (MAAR) if
        \begin{equation}\label{E:DMAAR}
                p(R=  r\mid Y =  y, \phi) =p(R= r\mid Y =  y', \phi),
            \end{equation}
        for all $\phi$, and all $ r,  y,  y'$ such that $\mathcal{S}( r, y) = \mathcal{S}( r, y')$.
    \end{definition}

    \begin{theorem}[\cite{bimetrkia2015Mealli}] \label{T:MAAR}
        Suppose that the missing data are MAAR, and:
        \begin{enumerate}
            \item the rows of $(Y,R)$ are exchangeable,
            \item the columns of $R$ are mutually conditionally independent given $Y$, 
            \item the probability that the $j^\text{th}$ variable is missing is positive for some $\phi$, $p(R_{i,j}=0\mid Y_{i \cdot}=y_{i \cdot},\phi)>0$, and moreover this probability depends on the $k^\text{th}$ component of $Y_{i \cdot}$, $Y_{i,k}$,
        \end{enumerate}
        \begin{equation}\label{E:Thrm1}
            p(R_{i,j}=0\mid Y_{i \cdot}=y_{i \cdot},\phi)\ne p(R_{i,j}=0\mid Y_{i,-k}=y_{i,-k},\phi),
        \end{equation}
        where the subscript``$-k$'' indicates the removal of $k^\text{th}$ element of the vector $Y_{i \cdot}$. Then $Y_{i,k}$ must be always observed for all $i$
        \[
            p(R_{i,k}=0\mid Y_{i \cdot}=y_{i \cdot},\phi)=0 \quad \text{for all } i \text{ and all } \phi. 
        \]
    \end{theorem}
        Theorem \ref{T:MAAR} implies that, if the vectors $Y_{\cdot j}$ and $Y_{\cdot k}$ have missing values, then both $R_{\cdot j}$ and $R_{\cdot k}$ can not depend on $Y_{\cdot k}$ or $Y_{\cdot j}$ given $\phi$ and the fully observed outcomes, whenever $J^*<J$. Below we formally state this corollary, the proof is given in Appendix 1.

    \begin{corollary}\label{C:main}
        Assume that only the first $J^*<J$ columns of $Y$ have a positive probability of having missing values, and the other $J - J^*$ columns of $Y$ are always fully observed. Under the conditions of Theorem \ref{T:MAAR},
        the probability that the $j^\text{th}$ variable is missing must only depend on the $J - J^*$ fully observed variables $Y_{\cdot J^*+1:J}$,
        \begin{equation}
            p(R_{i,j} = 0 \mid  Y_{i \cdot} = y_{i \cdot}, \phi) = p(R_{i,j} = 0 \mid  Y_{i, J^*+1:J} = y_{i, J^*+1:J}, \phi) 
        \end{equation}
        for all $\phi$, $i=1,\dots, N$ and, $j=1,\dots, J^*$.
    \end{corollary}
    By applying Bayes theorem, we can further show that conditional distribution of $Y_{\cdot j}$ is independent of $R_{\cdot j'}$ for $j,j' \in \{1,\dots, J^*\}$, given the observed variables $Y_{\cdot J^*+1:J}$. Below we formally state this corollary, the proof is given in Appendix 1.
    \begin{corollary}\label{C:small}
        Under the conditions of Corollary \ref{C:main} with parameters $\theta$ and $\phi$ distinct,
        \begin{equation} 
            p(Y_{\cdot j} =  y_{\cdot j} \mid  Y_{\cdot J^*+1:J} = y_{\cdot J^*+1:J}, R_{\cdot j'} = r_{\cdot j'}, \theta ) = 
            p(Y_{\cdot j} =  y_{\cdot j} \mid  Y_{\cdot J^*+1:J} = y_{\cdot J^*+1:J},\theta ).
        \end{equation}
    \end{corollary}




\section{Diagnostic tests for MAAR}\label{sec:Methods}

    In this section, we describe three diagnostic tests for detecting violations of Corollary \ref{C:main} and \ref{C:small} by introducing each in a simple case and then explaining how it extends to more general circumstances. Henceforth, we assume that the rows of $Y$ are independent and identically distributed, which simplifies our exposition but is not always necessary. 

    The diagnostic tests are formulated as hypothesis tests to allow for straightforward examination of their properties through a simulation study. Practitioners should use these diagnostic tests to identify problematic variables and conduct focused sensitivity analysis, rather than as formal accept-reject hypothesis tests. When one of our diagnostic tests ``rejects,'' we can conclude that some of the assumptions are likely to be violated; of course, we can not distinguish between a rejection due to a violation of the MAAR assumption as opposed to any of the other conditions, except by using domain knowledge.

\subsection{A comparison of conditional means approach}

    Consider three variables; $Y_{\cdot 1}$ and $Y_{\cdot 2}$ can have missing values, and $Y_{\cdot 3}$ is always fully observed. Under the conditions of Corollary \ref{C:small}, the conditional expectation of $Y_{\cdot 1}$ given $Y_{\cdot 3}$ is the same for the two partitions induced by $R_{\cdot 2}$,
    \begin{equation}\label{E:cma}
        E[Y_{i,1}\mid  Y_{i,3}=y_{i,3}, R_{i,2} = 0] = E[Y_{i,1}\mid  Y_{i,3}=y_{i,3}, R_{i,2} = 1] = E[Y_{i,1}\mid  Y_{i,3}=y_{i,3}].
    \end{equation}
    Therefore, $\mathcal{S}(R_{\cdot 2}, Y_{\cdot 1})$ and $\mathcal{S}(1-R_{\cdot 2}, Y_{\cdot 1})$ should have the same conditional distributions, given $Y_{\cdot 3}$. We can not directly test for a difference in $\mathcal{S}(R_{\cdot 2}, Y_{\cdot 1})$ and $\mathcal{S}(1-R_{\cdot 2}, Y_{\cdot 1})$ because we have not accounted for possible differences due to $Y_{\cdot 3}$. But, we can regress $Y_{\cdot 1}$ on $Y_{\cdot 3}$ and compare it to the regression of $Y_{\cdot 1}$ on $Y_{\cdot 3} \times R_{\cdot 2}$, for example using a likelihood ratio test. This is relatively straightforward; for example, if $Y_{\cdot 1}$ is Gaussian, then the maximum likelihood estimates are obtained by applying least squares regression on the $R_{\cdot 1}$ fully observed variables.
    The rejection of the smaller model in favor of the larger model, which includes the interaction of $R_{\cdot 2}$, implies that the assumptions of Corollary \ref{C:small} do not hold. 

    Similarly, $\mathcal{S}(R_{\cdot 1}, Y_{\cdot 2})$ and $\mathcal{S}(1-R_{\cdot 1}, Y_{\cdot 2})$ conditional on $Y_{\cdot 3}$, should also have the same distributions, and we can perform the analogous test. In this example, we have two hypotheses to test; to obtain valid $p$-values, we need to account for the multiple testing. 

    This diagnostic test does not work whenever two variables have the same missingness pattern, but if that is the case, it seems unlikely that the assumption that the columns of $R$ are mutually conditionally independent given $Y$ holds.

    The above diagnostic test extends to general situations where we have $J$ variables, $J^*$ of which have missing values. As $J^*$ grows, the number of hypothesis tests to be conducted grows exponentially. 


\subsection{Directly testing a postulated missingness mechanism approach}
    With three variables, $Y_{\cdot 1}$ and $Y_{\cdot 2}$ that can potentially have missing values, and $Y_{\cdot 3}$ that is always fully observed, by Corollary \ref{C:main}, a MAAR missingness mechanism can only be a function of the fully observed variable, $Y_{\cdot 3}$:
    \begin{equation}\label{E:Gen_Post}
        p(R_{\cdot 1} = r_{\cdot 1}\mid Y = y, \phi)= f_{1}^\phi(y_{\cdot 3}) \quad \text{and} \quad p(R_{\cdot 2} = r_{\cdot 2}\mid Y = y, \phi)= f_2^\phi(y_{\cdot 3}),
    \end{equation}
    for some functions $f_{1}^\phi$ and $f_2^\phi$. With specific forms for $f_{1}^\phi$ and $f_2^\phi$, we can directly test (\ref{E:Gen_Post}). For example, focusing on $R_{\cdot 1}$, assume that the missingness mechanism follows a logistic regression linear in $Y_{\cdot 3}$, and we test,
    \begin{align}\label{E:Logistic_Test}
        &H_0: \ p(R_{\cdot 1} = r_{\cdot 1}\mid Y = \tilde y, \phi)= \text{logit}^{-1}(\alpha+\beta_3 \tilde y_{\cdot 3}) \notag \\  \text{versus} \quad 
        &H_A: \ p(R_{\cdot 1} = r_{\cdot 1}\mid  Y = \tilde y, \phi) = \text{logit}^{-1}MAAR(\alpha + \beta_1\tilde y_{\cdot 1} + \beta_2 \tilde y_{\cdot 2} + \beta_3 \tilde y_{\cdot 3}).
    \end{align}
    ``Rejection" of $H_0$ implies that the assumptions of Corollary \ref{C:main} are violated. 

    Performing such a hypothesis test, with a specified missingness mechanism, can be challenging. One possible direction is to use multiple imputation \citep{rubin2004multiple} to generate imputations under the null hypothesis. Using the completed data sets, we can conduct the hypothesis test, specified in (\ref{E:Logistic_Test}), and obtain a $p$-value, using the appropriate adjustment as described in \cite{meng1992performing}. Once we have tested $R_{\cdot 1}$, we could also test a, possibly different, postulated missingness mechanism for $R_{\cdot 2}$. This diagnostic tests generalizes easily to any value of $J$ and $J^*$ as well as to any postulated missingness mechanism. 

\subsection{Gaussian copula approach}

    Suppose we model the joint distribution of $(Y, R)$ using a Gaussian copula, which implies a simple procedure that can be used both for diagnostic purposes and for generating multiple imputations \citep{hoff2007extending,Hollenbach2017cop}. Copulas factor the joint distribution of a multivariate variable into the univariate marginal distributions times a factor called the copula. This approach was initially developed in \citet{sklar1959fonctions}; see \citet{nelson:2006} for a more recent treatment. 

    The main benefit of the copula approach is the decoupling of the correlation from each of the marginal distribution, which can allow for simple modeling of continuous and categorical variables. The appeal of a semiparametric methods for copula models is that they treat the marginal distributions as nuisance parameters, and thereby reduce the amount of information that must be specified \emph{apriori}. In particular, we propose using the approach of \citet{hoff2007extending}, which employs the rank likelihood for semiparametric copula estimation.

    For unit $i$, define $W_{i \cdot}=(Y_{i \cdot},R_{i \cdot})= (Y_{i,1},\dots,Y_{i,J},R_{i,1},\dots R_{i,J^*})$, and assume
    \[
        W_{i,j} = F_{j}(\Phi(Z_{i,j})) \quad Z_{i,j} \sim N(0,C),\ j=1,\dots J, J+1,\dots,J+J^*,
    \] 
    where $F_{j}$ is the univariate CDF of variable j, $\Phi(x)$ is the CDF of the standard Gaussian evaluated at $x$, and $C$ is the correlation matrix, which we assume has an inverse-Wishart prior distribution. Details of the algorithm for estimation are in \citet{hoff2007extending}. 

    Once we obtain a posterior distribution of $C$, we can obtain the posterior distribution of $\mathrm{cor}(W_{i \cdot},W_{i'.}\mid W_{\mathcal{K}.})$ for any $i,i'$ and $\mathcal{K} \subset \{1,\dots,J+J^*\}$, which we use to test the hypothesis that $Y_{\cdot j}$ is correlated with $R_{\cdot j'}$ conditional on $Y_{\cdot J^*+1:J}$. If we reject this hypothesis, meaning that the variables are dependent in the latent scale, then we can conclude that the assumptions for Corollary \ref{C:main} are likely violated. 



\section{Simulation study}\label{sec:Simulation}

\subsection{Factors and their levels}
    We evaluate each of the three proposed diagnostic tests under an array of different scenarios by tracking the number of correct decisions made, where a decision is labeled correct when a procedure rejects an incorrect null hypothesis or fails to reject a correct null hypothesis. Our outcome of interest is the proportion of correct decisions made, as opposed to the type I error and power, because statisticians typically do not know the true missingness mechanism and are rarely interested in studying its properties; their primary aim is to conduct an appropriate analysis under explicitly stated plausible assumptions. 

    Suppose $Y$ follows a multivariate normal distribution with mean 0 and covariance matrix $\Sigma= \rho \mathbbm{1}_J^T\mathbbm{1}_J+(1-\rho)I_J$, where $\mathbbm{1}_J$ is a vector of length $J$ with entries all equal to 1, and $I_J$ is the $J\times J$ identity matrix, where $J=5$; $Y_{\cdot 1}, Y_{\cdot 2}$ and $Y_{\cdot 3}$ can potentially have missing values, whereas $Y_{.4}$ and $Y_{.5}$ are fully observed.

    We partition the five factors of our simulation study into three categories: nature's factors (uncontrolled by the investigator); nature's estimable factors (unknown to the investigator but estimable from the data) and; factors over which the investigator has direct control.

    Nature's unknown factors:
    \begin{itemize}
        \item [Factor 1: ] the missingness mechanism with levels: a MAAR mechanism that satisfies the conditions of Theorem \ref{T:MAAR} (denoted by MAAR), a MAAR mechanism where the columns of $R$ are not mutually conditionally independent given $Y$ (denoted by MAAR2), and a missing always not at random mechanism that satisfies the other conditions of Theorem \ref{T:MAAR} (denoted by MNAAR). Table \ref{tab:mechanism}, provides the specifications of each of the missingness mechanisms. 
    \end{itemize}
    Nature's known factors:
    \begin{itemize}
        \item [Factor 2: ] sample size, $N= 100, 500, 1000$;
        \item [Factor 3: ] average proportion of missing values per variable, $m= 0.2, 0.4, 0.6$.
    \end{itemize}
    Nature's estimable factors:
    \begin{itemize}
        \item [Factor 4: ] correlation between the columns of $Y$, $\rho= 0.2, 0.4, 0.6, 0.8$.
    \end{itemize}
    Factors under control of the investigator:
    \begin{itemize}
        \item [Factor 5: ] diagnostic test: comparison of conditional means approach (CCM), directly testing a postulated missingness mechanism (DTPMM), Gaussian copula approach (GC).
    \end{itemize}

\begin{table}[]
\centering
\caption{The dependence of the missingness mechanism for the simulation study. For example, the first entry in the $R_{i, 1}$ column means that $p(R_{i,1} = 1\mid  Y=y, R_{i,-1}=r_{i,-1} ,\phi) = \text{logit}^{-1}(\alpha + y_{i,4} - y_{i,5})$. The constant $\alpha$ is used to determine the average proportion of missing values per variable.}
\setlength{\tabcolsep}{3pt}
\label{tab:mechanism}
\small{
\begin{tabular}{crrr}
\hline \hline 
Mechanism &$R_{i,1}$ & $R_{i,2}$ & $R_{i,3}$ \\ \hline
MAAR & $y_{i,4} - y_{i,5}$ & $y_{i,4} - y_{i,5}$ & $y_{i,4} - y_{i,5}$ \\
MAAR2  & $y_{i,4} - y_{i,5}$ & $y_{i,1}r_{i,1} + y_{i,4} - y_{i,5}$ &
 $y_{i,1}r_{i,1} +y_{i,2}r_{i,2}  + y_{i,4} - y_{i,5}$ \\
 MNAAR & $y_{i,4} - y_{i,5}$ & $\frac{1}{2}y_{i,1}+  \frac{1}{2}y_{i,3} + y_{i,4} - y_{i,5}$ &
 $y_{i,1} +y_{i,2}  + y_{i,4} - y_{i,5}$ \\
\hline  
\end{tabular}
}
\end{table}

\subsection{Results}

    For each possible combination of the factors, we generated 1,000 independent replications. Table \ref{tab:ANOVA} in Appendix 2 presents an analysis of variance of the 5-factor study to suggest which tables we should examine. The major sources of variation are the sample size $N$, Factor 2, followed by the missingness mechanism, Factor 1, and the diagnostic test used, Factor 5. Because the missingness mechanism is unknown, we focus the discussion of the results on the behavior of the diagnostic tests under nature's known and estimable factors. An analysis of the results across different levels of the missingness mechanism factor is included in Appendix 2. 

     Table \ref{tab:N_res} shows the proportion of correct decisions made by the three diagnostic tests for different sample sizes, correlations $\rho$, and proportion of missing values per variable ($m$). For large sample sizes, all three diagnostic tests reach the correct decision over 97\% of the time, for all levels of the other factors. For small samples sizes, directly testing a postulated missingness mechanism performs poorly because the likelihood ratio test used is based on the asymptotic distribution of the test statistic and has low power in small samples \citep{liu2017evaluation}. In small samples, the Gaussian copula slightly outperforms the comparison of conditional means approach, but the difference becomes negligible as the sample size increases.

\begin{table}[]
\setlength{\tabcolsep}{3pt}
\centering
\caption{Correct decision rates average over the correlation and missingness coefficient factors.}
\label{tab:N_res}
\small{\begin{tabular}{lllrrrlrrrlrrrlrrrlr}
\hline \hline 
\multicolumn{1}{c}{\multirow{2}{*}{\begin{tabular}[c]{@{}c@{}}Diagnostic Test\end{tabular}}} & \multicolumn{1}{l}{\multirow{2}{*}{\begin{tabular}[l]{@{}c@{}}Sample\\ Size\end{tabular}}} &  & \multicolumn{3}{c}{$\rho = 0.2$} &  & \multicolumn{3}{c}{$\rho = 0.4$} &  & \multicolumn{3}{c}{$\rho = 0.6$} &  & \multicolumn{3}{c}{$\rho = 0.8$} && \multicolumn{1}{c}{\multirow{2}{*}{\begin{tabular}[c]{@{}c@{}}Average\end{tabular}}} \\ \cline{4-6} \cline{8-10} \cline{12-14} \cline{16-18} 
\multicolumn{1}{c}{} & \multicolumn{1}{c}{} & \multicolumn{1}{r}{$m=$} & \multicolumn{1}{c}{0.2} & \multicolumn{1}{c}{0.4} & \multicolumn{1}{c}{0.6} &  & \multicolumn{1}{c}{0.2} & \multicolumn{1}{c}{0.4} & \multicolumn{1}{c}{0.6} &  & \multicolumn{1}{c}{0.2} & \multicolumn{1}{c}{0.4} & \multicolumn{1}{c}{0.6} &  & \multicolumn{1}{c}{0.2} & \multicolumn{1}{c}{0.4} & \multicolumn{1}{c}{0.6} & & \\ \hline
\multirow{3}{*}{CCM} 
 & 100 & & .70    & .74    & .61    &  & .69    & .71    & .60    &  & .60    & .62    & .52    &  & .47    & .49    & .44   && .60 \\
    & 500  && .98    & .99    & .99    &  & .98    & .98    & .98    &  & .98    & .98    & .98    &  & .97    & .98    & .94   && .98 \\
    & 1000 && .98    & .99    & .98    &  & .98    & .98    & .98    &  & .98    & .98    & .98    &  & .99    & .98    & .99 && .98 \\
 &  &  & \multicolumn{1}{l}{} & \multicolumn{1}{l}{} & \multicolumn{1}{l}{} &  & \multicolumn{1}{l}{} & \multicolumn{1}{l}{} & \multicolumn{1}{l}{} &  & \multicolumn{1}{l}{} & \multicolumn{1}{l}{} & \multicolumn{1}{l}{} &  & \multicolumn{1}{l}{} & \multicolumn{1}{l}{} & \multicolumn{1}{l}{} \\
\multirow{3}{*}{DTPMM} 
 & 100 &  & .38 & .33 & .33 &  & .36 & .33 & .33 &  & .35 & .33 & .33 &  & .34& .33 & .33 & &.34\\
 & 500 &  & .99 & .99 & .80 &  & .99 & .98 & .74 &  & .98 & .93 & .56 &  & .82& .65 & .37 & &.82\\
 & 1000 &  & 1 & 1 & 1 &  & 1 & 1 & .99 &  & 1 & 1 & .98 &  & .99 & .97 & .69  & &.97\\
 &  &  & \multicolumn{1}{l}{} & \multicolumn{1}{l}{} & \multicolumn{1}{l}{} &  & \multicolumn{1}{l}{} & \multicolumn{1}{l}{} & \multicolumn{1}{l}{} &  & \multicolumn{1}{l}{} & \multicolumn{1}{l}{} & \multicolumn{1}{l}{} &  & \multicolumn{1}{l}{} & \multicolumn{1}{l}{} & \multicolumn{1}{l}{} \\
\multirow{3}{*}{GC} 
 & 100 &  & .78 & .83 & .68 &  & .79 & .82 & .68 &  & .73 & .76 & .63 &  & .58 & .60 & .52  & & .70 \\
 & 500 &  & .98 & .98 & .98 &  & .98 & .98 & .98 &  & .98 & .98 & .99 &  & .98 & .99 & .94  & & .98 \\
 & 1000 &  & .98 & .99 & .98 &  & .98 & .98 & .98 &  & .98& .98 & .98 &  & .98 & .98 & .99  & & .98 \\\hline
\end{tabular}}
\end{table}

    The main advantage of running diagnostics individually for each of the missingness indicators is to identify variables that are likely to violate the assumptions of Theorem \ref{T:MAAR}. Table \ref{tab:identification} presents the correct decision rates for each of the missingness indicators as well as the overall rate for the three diagnostic tests across the different sample sizes. We present the results averaged over the correlation and proportion of missing values because, as reflected in Table \ref{tab:ANOVA} the effect of their interaction is small. Under all three missingness mechanisms, $R_{\cdot 1}$ always satisfies the conditions of Theorem \ref{T:MAAR}. For all sample sizes, all three diagnostic tests make the correct decisions regarding $R_{\cdot 1}$ at least 97\% of the time. Under the MAAR2 and MNAAR missingness mechanisms, both $R_{\cdot 2}$ and $R_{\cdot 3}$ violate the conditional independence assumption ($R_{\cdot 2}$ depends on $R_{\cdot 1}$, and $R_{\cdot 3}$ depends on both $R_{\cdot 1}$ and $R_{\cdot 2}$), and the MAAR mechanism assumption ($R_{\cdot 3}$ has a greater dependence than $R_{\cdot 2}$ on variables with missing values), respectively. As expected, the diagnostic tests have a higher correct decision rate for $R_{\cdot 3}$ relative to $R_{\cdot 2}$; however, the difference reduces as the sample size increases. Again, the comparison of conditional means and Gaussian copula diagnostic tests have relatively similar performance, and directly testing a postulated missingness mechanism does not perform well when the sample size are small.

\begin{table}[]

\setlength{\tabcolsep}{3pt}
\centering
\caption{Correct decision rates for each of the three missingness indicators averaged over the missingness mechanism, sample size and correlation.}
\label{tab:identification}
\begin{tabular}{llrrr} \hline \hline
Diagnostic Test & $N$ & $R_{\cdot 1}$ & $R_{\cdot 2}$ & $R_{\cdot 3}$ \\ \hline
    & 100  & .98 & .42 & .55 \\
CCM    & 500  & .98 & .85 & .98 \\
    & 1000 & .98 & .97 & .99 \\ \\
    & 100  & 1    & .33 & .34 \\
DTPMM    & 500  & 1    & .47 & .82 \\
    & 1000 & 1    & .74 & .97 \\\\
    & 100  & .99 & .47 & .66 \\ 
GC    & 500  & .98 & .91 & .98 \\
    & 1000 & .97 & .98 & .99 \\ \hline
\end{tabular}

\end{table}

\subsection{Discussion}

Our simulations suggest that statisticians should use either the Gaussian copula or the comparison of conditional means approach. The choice between the two diagnostic tests depends on the statistician's beliefs about the correlation structures in $Y$. Directly testing a postulated missingness mechanism should be avoided for small samples and only used if the statistician can provide a plausible model for the missingness mechanism. 

In this simulation study, we did not consider scenarios for which the global modeling assumptions are violated. We expect to see a reduction in performance due to model misspecification; however, we still believe that these tests can provide the statistician with useful information regarding which variables are likely to violate the MAAR missignness assumptions. 


\section{Conclusion}\label{sec:Conclusion}

Using Corollary \ref{C:main} and \ref{C:small}, we proposed three diagnostic tests for detecting the validity of the assumptions in Theorem 1, using the observed data alone. We showed, through simulation, that all three diagnostic tests had high discriminatory power and can identify the variables that violate the assumptions. As expected, we cannot distinguish between violations of the MAAR assumption and the assumption that the columns of $R$ are mutually conditionally independent given $Y$.

We encourage practitioners to use these diagnostic tests in the diagnosis stage to identify problematic variables and then conduct a targeted sensitivity analysis to assess the impact of violations of suspect assumptions on the scientific conclusions. The identification of problematic variables is especially important when the number of variables is large because most sensitivity analysis methods are computationally intensive. 

\bibliographystyle{apalike}
\bibliography{reference}

\section*{Appendix 1}
\subsection*{Proofs of Corollary 1 and 2}

  \begin{proof}[of Corollary 1]
        By contradiction. Suppose that,
        \[
            p(R_{i,j} = 0 \mid  Y_{i \cdot} = y_{i \cdot}, \phi) \ne p(R_{i,j} = 0 \mid  Y_{i, J^*+1:J} = y_{i, J^*+1:J}, \phi).
        \]
        Then, there must exist a set $S_j \subset\{1,\dots, J^*\}$ such that
        \[
            p(R_{i,j} = 0 \mid  Y_{i \cdot} = y_{i \cdot}, \phi) = p(R_{i,j} = 0 \mid  Y_{i, S_j \cup J^*+1:J} = y_{i, S_j \cup J^*+1:J}, \phi),
        \]
        and removing any element from $S_j$ breaks the equality. Theorem \ref{T:MAAR} implies that for all $k\in S_j$, 
        \[
            p(R_{i,k}=0\mid Y_{i \cdot}=y_{i \cdot},\phi) = 0 \quad \text{for all } i \text{ and all } \phi,
        \]
        which contradicts the assumption that the $k^\text{th}$ column of $Y$ has positive probability of having missing values.
    \end{proof}

\begin{proof}[of Corollary 2]
    By Bayes theorem,
    \begin{align*}
        &p(Y_{\cdot j} = y_{\cdot j} \mid  Y_{\cdot J^*+1:J} = y_{\cdot J^*+1:J}, R_{\cdot j'} = r_{\cdot j'}, \theta ) \\
        &= \frac{p(Y_{\cdot j} =  y_{\cdot j} \mid  Y_{\cdot J^*+r1:J} = y_{\cdot J^*+1:J}, \theta )p(R_{\cdot j'} = r_{\cdot j'}\mid Y_{\cdot j} =  y_{\cdot j}, Y_{\cdot J^*+1:J} = y_{\cdot J^*+1:J})}{p(R_{\cdot j'} = r_{\cdot j'}\mid Y_{\cdot J^*+1:J} = y_{\cdot J^*+1:J})} \\
        &= \frac{p(Y_{\cdot j} =  y_{\cdot j} \mid  Y_{\cdot J^*+r1:J} = y_{\cdot J^*+1:J}, \theta )\int p(R_{\cdot j'} = r_{\cdot j'}\mid Y_{\cdot j} =  y_{\cdot j}, Y_{\cdot J^*+1:J} = y_{\cdot J^*+1:J},\phi) p(\phi) d\phi}{p(R_{\cdot j'} = r_{\cdot j'}\mid Y_{\cdot J^*+1:J} = y_{\cdot J^*+1:J})} \\
        &= \frac{p(Y_{\cdot j} =  y_{\cdot j} \mid  Y_{\cdot J^*+r1:J} = y_{\cdot J^*+1:J}, \theta )\int p(R_{\cdot j'} = r_{\cdot j'}\mid Y_{\cdot J^*+1:J} = y_{\cdot J^*+1:J},\phi) p(\phi) d\phi}{p(R_{\cdot j'} = r_{\cdot j'}\mid Y_{\cdot J^*+1:J} = y_{\cdot J^*+1:J})} \\
        &= p(Y_{\cdot j} =  y_{\cdot j} \mid  Y_{\cdot J^*+1:J} = y_{\cdot J^*+1:J}, \theta ).
    \end{align*}
    The third equality holds by Corollary \ref{C:main}. 
\end{proof}

\section*{Appendix 2}
\subsection*{Further discussion of the results from the simulation study}

\begin{table}[]
\centering
\caption{Analysis of Variance of the 5-Factor Simulation Study}
\setlength{\tabcolsep}{3pt}
\label{tab:ANOVA}
\begin{tabular}{lrr}
\hline \hline 
\multirow{2}{*}{Source} & \multicolumn{1}{c}{\multirow{2}{*}{\begin{tabular}[c]{@{}c@{}}Degrees of\\ Freedom\end{tabular}}} & \multirow{2}{*}{Mean Square $\times 10^4$}\\  \\
\hline  
Diagnostic Test                         & 2 &  9465 \\ 
$N$                             &  2 & 60044 \\ 
$\rho$                            & 3 &  1657 \\ 
$m$                      & 2 &  1556 \\ 
Mechanism                        & 2 &  9187 \\ 
Diagnostic Test $\times$ $N$                       & 4 &  2906 \\ 
Diagnostic Test $\times$ $\rho$                     & 6 &  108 \\ 
$N$ $\times$ $\rho$                          & 6 &  183 \\ 
Diagnostic Test $\times$ $m$               & 4 &  428 \\ 
$N$ $\times$ $m$                    & 4 &  220 \\ 
$\rho$ $\times$ $m$                  & 6 &  42 \\ 
Diagnostic Test $\times$ Mechanism                 & 4 &  3766 \\ 
$N$ $\times$ Mechanism                      & 4 &  5559 \\ 
$\rho$ $\times$ Mechanism                    & 6 &  490 \\ 
$m$ $\times$ Mechanism              & 4 &  432 \\ 
Diagnostic Test $\times$ $N$ $\times$ $\rho$                   &12 &  394 \\ 
Diagnostic Test $\times$ $N$ $\times$ $m$             & 8 &  481 \\ 
Diagnostic Test $\times$ $\rho$ $\times$ $m$           &12 &  54 \\ 
$N$ $\times$ $\rho$ $\times$ $m$                &12 &  48 \\ 
Diagnostic Test $\times$ $N$ $\times$ Mechanism               & 8 &  644 \\ 
Diagnostic Test $\times$ $\rho$ $\times$ Mechanism             &12 &  33 \\ 
$N$ $\times$ $\rho$ $\times$ Mechanism                  &12 &  47 \\ 
Diagnostic Test $\times$ $m$ $\times$ Mechanism       & 8 &  103 \\ 
$N$ $\times$ $m$ $\times$ Mechanism            & 8 &  67 \\ 
$\rho$ $\times$ $m$ $\times$ Mechanism          &12 &  13 \\ 
Diagnostic Test $\times$ $N$ $\times$ $\rho$ $\times$ $m$         &24 &  29 \\ 
Diagnostic Test $\times$ $N$ $\times$ $\rho$ $\times$ Mechanism           &24 &  104 \\ 
Diagnostic Test $\times$ $N$ $\times$ $m$ $\times$ Mechanism     &16 &  120 \\ 
Diagnostic Test $\times$ $\rho$ $\times$ $m$ $\times$ Mechanism   &24 &  14 \\ 
$N$ $\times$ $\rho$ $\times$ $m$ $\times$ Mechanism        &24 &  17 \\ 
Diagnostic Test $\times$ $N$ $\times$ $\rho$ $\times$ $m$ $\times$ Mechanism &48 &  11 \\ 
\hline  
\end{tabular}
\end{table}

\begin{table}[]
\setlength{\tabcolsep}{3pt}
\centering
\caption{Correct choice rate averaged over the correlation and missingness factors. }
\label{tab:missing_res}
\small{\begin{tabular}{lllrrrlrrrlrrrlrrr}
\hline \hline 
\multicolumn{1}{c}{\multirow{2}{*}{\begin{tabular}[c]{@{}c@{}}Missingness\\ Mechanism\end{tabular}}} & \multicolumn{1}{c}{\multirow{2}{*}{\begin{tabular}[c]{@{}c@{}}Diagnostic\\ Test\end{tabular}}} &  & \multicolumn{3}{c}{$R_{\cdot 1}$} &  & \multicolumn{3}{c}{$R_{\cdot 2}$} &  & \multicolumn{3}{c}{$R_{\cdot 3}$} &  & \multicolumn{3}{c}{Overall} \\ \cline{4-6} \cline{8-10} \cline{12-14} \cline{16-18} 
\multicolumn{1}{c}{} & \multicolumn{1}{c}{} & \multicolumn{1}{r}{$N=$} & \multicolumn{1}{c}{100} & \multicolumn{1}{c}{500} & \multicolumn{1}{c}{1000} &  & \multicolumn{1}{c}{100} & \multicolumn{1}{c}{500} & \multicolumn{1}{c}{1000} &  & \multicolumn{1}{c}{100} & \multicolumn{1}{c}{500} & \multicolumn{1}{c}{1000} &  & \multicolumn{1}{c}{100} & \multicolumn{1}{c}{500} & \multicolumn{1}{c}{1000} \\ \hline
\multirow{3}{*}{MAAR} 
  & CCM &  & .98 & .99 & .98 &  & .99 & .99 & .99 &  & .98 & .98 & .98 &  & .95 & .96 & .95 \\
 & DTPMM &  & 1 & 1 & 1 &  & 1 & 1 & 1 &  & 1 & 1 & 1 &  & 1 & 1 & 1 \\
 & GC &  & .99 & .99 & .98 &  & .99 & .99 & .98 &  & .99 & .99 & .98 &  & .98 & .96 & .95 \\
 &  &  & \multicolumn{1}{l}{} & \multicolumn{1}{l}{} & \multicolumn{1}{l}{} &  & \multicolumn{1}{l}{} & \multicolumn{1}{l}{} & \multicolumn{1}{l}{} &  & \multicolumn{1}{l}{} & \multicolumn{1}{l}{} & \multicolumn{1}{l}{} &  & \multicolumn{1}{l}{} & \multicolumn{1}{l}{} & \multicolumn{1}{l}{} \\
\multirow{3}{*}{MAAR2} 
 & CCM &  & .98 & .98 & .98 &  & .17 & .88 & .99 &  & .32 & .98 & 1 &  & .42 & .99 & 1 \\
 & DTPMM &  & 1 & 1 & 1 &  & 0 & .24 & .66 &  & .01 & .68 & .94 &  & .01 & .69 & .94 \\
 & GC &  & .98 & .97 & .95 &  & .28 & .94 & 1 &  & .48 & .98 & 1 &  & .57 & .99 & 1 \\
 &  &  & \multicolumn{1}{l}{} & \multicolumn{1}{l}{} & \multicolumn{1}{l}{} &  & \multicolumn{1}{l}{} & \multicolumn{1}{l}{} & \multicolumn{1}{l}{} &  & \multicolumn{1}{l}{} & \multicolumn{1}{l}{} & \multicolumn{1}{l}{} &  & \multicolumn{1}{l}{} & \multicolumn{1}{l}{} & \multicolumn{1}{l}{} \\
\multirow{3}{*}{MNAAR} 
 & CCM &  & .98 & .98 & .98 &  & .1 & .68 & .94 &  & .36 & .99 & 1 &  & .42 & .99 & 1 \\
 & DTPMM &  & 1 & 1 & 1 &  & 0 & .16 & .55 &  & .02 & .77 & .97 &  & .02 & .77 & .97 \\
 & GC &  & .99 & .98 & .98 &  & .13 & .78 & .96 &  & .49 & .99 & 1 &  & .55 & .99 & 1 \\\hline
\end{tabular}}
\end{table}

Table \ref{tab:missing_res} shows the correct decision rates for $R_{\cdot 1},R_{\cdot 2}$ and $R_{\cdot 3}$ as well as the overall rate, for different sample sizes and the three different missingness mechanisms. Under the MAAR mechanism, directly testing a postulated missingness mechanism approach never rejected the null hypothesis; \cite{liu2017evaluation} similarly showed that the likelihood ratio test has low empirical Type I error for small samples with a significant proportion of missing data. The Gaussian copula and comparison of conditional means approaches had rejection rates close to the specified nominal level. For low values of $N$, the Gaussian copula and the comparison of conditional means approaches had higher rejection rates for the MNAAR and MAAR2 mechanisms than directly testing a postulated missingness mechanism; however, as the sample size increased, the difference decreased. 

All three diagnostic tests were able to identify the variables that were violating the MAAR/conditional independence assumption, although they needed a larger sample size to detect violations in $R_{\cdot 2}$ relative to $R_{\cdot 3}$. The correct decision rates under the MAAR2 and MNAAR missingness mechanisms are similar because there is no information in the data that distinguishes between them.

\end{document}